\documentclass{PoS}

\usepackage{graphicx}
\usepackage{multirow}

\newlength{\photoheight}
\newlength{\photheight}
\def\rmsub#1#2{#1_{\mbox{\tiny #2}}}	    
\def\rmsup#1#2{#1^{\mbox{\tiny #2}}}	    

\def\nth{\rmsup{n}{th}}

\def\sec{\(\S\)}

\def\naive{na{\"{\i}ve}}

\def\phil{\rmsub{\phi}{l}}
\def\phis{\rmsub{\phi}{s}}

\def\Sg{\rmsub{S}{g}}
\def\Sf{\rmsub{S}{f}}
\def\Seff{\rmsub{S}{eff}}

\def\ratdeg{n}
\def\polydeg{\ratdeg}

\def\exp{\mathop{\rm exp}}		    
\def\tr{\mathop{\rm tr}}		    
\def\ln{\mathop{\rm ln}}		    
\def\det{\mathop{\rm det}}		    

\def\M{{\cal M}}			    
\def\D{{\mathcal{D}}}
\def\dt{\delta\tau}			    
\def\dH{\delta H}			    
\def\pacc{P_{\mbox{\tiny acc}}}		    

\def\m{\M}

\def\tint#1{\tau_{\mbox{\tiny{int}}}^{\mbox{\tiny{#1}}}}

\def\Nmv{\rmsub{N}{mv}}

\def\Niter{\rmsub{N}{iter}}
\def\Nf{\rmsub{N}{f}}
\def\Sf{\rmsub{S}{f}}

\def\mf#1{M_{\mbox{\tiny{#1}}}^{\dagger}M_{\mbox{\tiny{#1}}}}

\def\Ml{\rmsub{\m}{l}}
\def\Ms{\rmsub{\m}{s}}

\def\ml{\rmsub{m}{l}}
\def\ms{\rmsub{m}{s}}

\def\asqtad{AsqTad}

\title{The Rational Hybrid Monte Carlo Algorithm}

\ShortTitle{The Rational Hybrid Monte Carlo Algorithm}

\author{\speaker{M. A. Clark}\\ Center for Computational Science,
        Boston University, Boston, MA 02215, United States of
        America\\ E-mail: \email{mikec@bu.edu}}

\abstract{The past few years have seen considerable progress in
algorithmic development for the generation of gauge fields including
the effects of dynamical fermions.  The Rational Hybrid Monte Carlo
(RHMC) algorithm, where Hybrid Monte Carlo is performed using a
rational approximation in place the usual inverse quark matrix kernel
is one of these developments.  This algorithm has been found to be
extremely beneficial in many areas of lattice QCD (chiral fermions,
finite temperature, Wilson fermions etc.).  We review the algorithm
and some of these benefits, and we compare against other recent
algorithm developements.  We conclude with an update of the Berlin
wall plot comparing costs of all popular fermion formulations.}

\FullConference{XXIV International Symposium on Lattice Field Theory\\
		 July 23-28 2006\\
		 Tucson Arizona, US}

\begin{document}

\section{Introduction}

\noindent We start as always with the Lattice QCD path integral
\[
\langle\Omega\rangle
= {1\over Z} \int [dU] e^{-\Sg(U)} [\det\M(U)]^\alpha \Omega(U)\,
\label{eq:fi}
\]
where \(\alpha=\frac{\Nf}{4}\) (\(\frac{\Nf}{2}\)) for staggered
(Wilson) fermions, \(\m=M^\dagger M\) with \(M\) the discretised Dirac
operator.  This implies we can define a theory with an arbitrary
number of fermion flavours \(\Nf\) if we are willing to allow a
non-integer \(\alpha\) parameter.  In this regime the conventional
Hybrid Monte Carlo (HMC) algorithm fails because there is no method
where by which we can directly evaluate either the action or its
variation with respect to the gauge field to evaluate the force.
Since current dynamical QCD calculations are focused on \(\Nf=2+1\) we
must use other algorithms if we are to proceed.

\section{Hybrid Algorithms}
\subsection{Hybrid Monte Carlo (HMC)}

\noindent Before we describe algorithms for non-integer \(\alpha\), we
describe the HMC algorithm \cite{Duane:1987de}, which is the {\it de
facto} algorithm for fermion theories where \(\alpha=1\).  Here we
rewrite the fermion determinant in terms of pseudo-fermions
\[
\det\M = \int D\phi^\dagger D\phi e^{-\phi^\dagger \m^{-1} \phi} =
\int D\phi^\dagger D\phi \, e^{-\Sf}.
\]
We introduce a fictitious momentum field and define a Hamiltonian \( H
= \frac{1}{2}\tr \pi^2 + \Sg + \Sf \).  With the Hamiltonian defined,
we can then evolve the gauge field through integrating Hamilton's
equations of motion: the resulting gauge fields will have the desired
probability distribution,
\[
P(U,\phi) = {1\over Z} e^{-\Sg -\Sf}.
\label{eq:prob}
\]
To ensure ergodicity of the algorithm, the momentum field (and
pseudofermion fields) must be refreshed periodically, once per
trajectory.  Integrating Hamilton's equations requires we discretise
the fictitious time dimension and use a numerical integration scheme
with stepsize \(\dt\) to evolve the gauge field (the so called
molecular dynamics (MD)).  This introduces an \(O(\dt^k)\) error to
the distribution, where \(k\) is equal to the order of integration
scheme used.  This error can be stochastically corrected for with the
use of a Metropolis acceptance test at the end of the each trajectory.
The Metropolis acceptance test requires detailed balance, this places
two constraints on the integration scheme chosen: it must be both
reversible and area preserving.  These constraints are maintained if
we use symmetric symplectic integrators, the most simple of these of
course being the second order leapfrog integrator.  Hence the HMC
algorithm is an exact algorithm and the results obtained are
independent of the stepsize, the stepsize is thus chosen to give a
non-negligible acceptance rate.

Each update of HMC consists of
  \begin{itemize}
  \item Hybrid Molecular Dynamics Trajectory
    \begin{itemize}
    \item Momentum refreshment heatbath (\(P(\pi) \propto
      e^{-\pi^*\pi/2}\)).
    \item Pseudo-fermion heatbath (\(\phi \propto M^\dagger\xi\), where
      \(P(\xi)\propto e^{-\xi^*\xi}\))
    \item MD trajectory with \(\tau/\delta\tau\) steps
    \end{itemize}
  \item Metropolis Acceptance Test \(\pacc = \mbox{min}(1,e^{-\dH})\).
  \end{itemize}
The cost of the HMC algorithm rapidly increases with decreasing
fermion mass; there are two principle sources for this mass
dependence.  We have to evaluate the inverse of the Dirac operator
applied to a vector for each force evaluation, this is typically done
using a Krylov solver such as conjugate gradient (CG), this cost blows
up as the fermion mass decreases because of associated the condition
number increase.  The other reason is because of increase in fermion
force magnitude which necessitates a decrease in stepsize to maintain
a constant acceptance rate.  The end result is that including the
effect of physical dynamical fermions is an unrealistic prospect.
Fortunately there have been improvements to the basic algorithm which
vastly reduce this cost: these shall be covered in \sec\ref{sec:multiple}.

\subsection{The $R$ Algorithm}

An alternative to using the pseudofermion formulation of HMC is
rewrite the fermionic determinant as an exponential trace log,
resulting in a fermion effective action
\[
\det\m^{\alpha} = \exp\left(\alpha\tr\ln \m\right) = \exp\left( -\Seff \right).
\]
The advantage of this method is that we can represent a theory with an
arbitrary \(\alpha\) parameter.  As before with the HMC algorithm, we
integrate Hamilton's equations to give us the desired probability
distribution of the gauge fields.  However, when it comes down to
evaluating the force, the presence of the trace in the action requires
that we evaluate the explicit matrix inverse.  This is obviously
unfeasible, so the trace is represented by a noisy estimator: the
resulting force term is equivalent to the pseudofermion force which is
present in HMC.

The problem with this method is that because a noisy estimator for the
fermion force has been used, a new \(O(\dt)\) error is introduced.
Through the introduction of an irreversible and non-area preserving
integration scheme an algorithm with \(O(\dt^2)\) leading error can
be obtained.  However, this breaks detailed balance so the
algorithm cannot be made exact through the inclusion of a Metropolis
acceptance test, hence the algorithm is inexact and any results
obtained from the algorithm will have a dependence on the integrating
stepsize.

The cost of the \(R\) algorithm is \naive ly equal to HMC in that one
linear equation solve is required per force evaluation, however,
strictly speaking an extrapolation to zero stepsize is required.  This
is an expensive and time consuming process.  For many years,
calculations done using staggered fermions with the \(R\) algorithm
have used the rule of thumb \(\dt^R \sim \frac{2}{3}\ml\); the
extrapolation to zero stepsize has been rarely carried out.  This thus
raises a potential question mark over any results generated since the
size of the stepsize error is unknown.

\subsection{Polynomial Hybrid Monte Carlo (PHMC)}

Like HMC upon which PHMC is based on, we start by representing the
determinant in terms of pseudofermions.  The kernel in the
pseudofermion bilinear is replaced by an optimal polynomial
approximation to the desired matrix function,
\[
  \det\m^{\alpha} = \int \D \phi^\dagger\D\phi
  e^{-\phi^\dagger\M^{-\alpha}\phi} \approx \int \D\phi^\dagger\D\phi
  e^{-\phi^\dagger P(\m)\phi},
\]
where \(P(\M)\) is valid over the spectral range of the operator
\cite{deForcrand:1996ck,Frezzotti:1997ym}.  The roots of optimal
polynomial approximations come in complex conjugate pairs,
\(P(\M)=p^\dagger(\m)p(\m)\); this means that the heatbath can easily
be evaluated, \(\phi = p(\m)^{-1} \eta\).\footnote{Presumably the
cheapest method to evaluate the heatbath is to expand the polynomial
inverse in terms of partial fractions and to evaluate using a
multi-shift solver.}

A polynomial approximation has equivalent convergence properties to a
Jacobi iteration.  Such methods are known to be inferior to Krylov
methods, hence there is potential for PHMC to be an expensive
algorithm compared to HMC (i.e., the polynomial degree \(\polydeg\)
will be much greater than \(\Niter\) the number of Krylov iterations).
One approach to alleviate this problem is to use a low degree
polynomial to represent the fermion action, and to correct for this by
reweighting either the acceptance test or any observables by the
required determinant ratio.  The problem with this approach is that
the difference between the guidance Hamiltonian and the actual
Hamiltonian is extensive in the volume; this results in a poorer
scaling with volume compared to HMC, however, there have been recent
improvements in the PHMC algorithm which minimise this effect
\cite{Scholz:2006hd}.

For theories with \(\alpha \ne 1\), the derivative must be taken using
the Leibniz rule, hence we have to sum over \(\polydeg\) terms, this
is costly with respect to memory (we need to simultaneously store
\(\polydeg/2\) vectors, with \(n\sim O(100-1000)\)) and there is also
increased potential for rounding errors.

\section{Rational Hybrid Monte Carlo (RHMC)}

\begin{figure}
  \begin{center}
    \includegraphics[height=5cm]{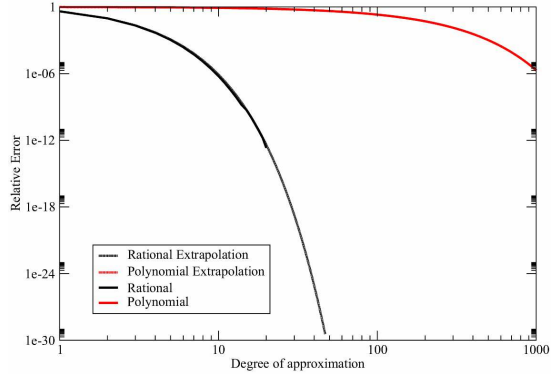}
    \caption{A comparison of the relative error between rational and
      polynomial approximation to the inverse square root function
      (spectral bounds = \([3\times10^{-3},1]\)).}
    \label{fig:degree}
  \end{center}
\end{figure}

An alternative to using a polynomial kernel is to use a rational
kernel.  In figure \ref{fig:degree} there is a comparison between
equivalent polynomial and rational approximations, plotting the
relative error as a function of approximation degree.  Rational
approximations have far superior convergence properties: over the
spectral bounds we are concerned with, we can achieve an approximation
to the desired function with maximum relative error that is good to
machine precision (\(\sim 10^{-15}\)) with approximation degree
\(O(20)\) or less (compared to \(O(1000)\) with polynomials).

Optimal rational approximations are generated using the Remez
algorithm \cite{Remez:1962}, the roots and poles of which are in
general real.  Thankfully the poles are also always positive for
\(|\alpha|<1\) which are the functions of interest.  When written in
partial fraction form (\(r(x) =
\sum_{k=1}^{m}\frac{\alpha_k}{x+\beta_k}\)) they can be evaluated
using a multi-shift solver \cite{Frommer:1995ik}: hence the cost
evaluating a rational function is essentially the same as a single
matrix inversion, and to first order the precision is independent of
the cost.  It is also worth mentioning that since the \(\alpha_k\)
coefficients are in general all the same sign (for \(|\alpha|<1\)) the
evaluation of rational functions using partial fractions is
numerically stable.

Exactly as was done with PHMC, we rewrite the determinant in terms of
pseudofermions, but now replace the kernel in the bilinear by a
rational approximation,
\[
  \det\m^{\alpha} = \int \D \phi^\dagger\D\phi
  e^{-\phi^\dagger\m^{-\alpha}\phi} \approx \int \D\phi^\dagger\D\phi
  e^{-\phi^\dagger r^2(\m)\phi},
\]
with \( r(x) = x^{-\alpha/2} \).  Unlike the case of PHMC, we have to
enforce the degeneracy condition with a rational kernel to allow the
heatbath to be evaluated.  The great advantage now though is because
we can include a rational approximation to arbitrary precision there
is no requirement to reweight the acceptance test \cite{Clark:2003na}.
Hence we can proceed as exactly as the conventional HMC algorithm.
The RHMC algorithm thus consists of
\begin{itemize}
\item Hybrid Molecular Dynamics Trajectory
  \begin{itemize}
  \item Momentum refreshment heatbath (\(P(\pi) \propto
    e^{-\pi^*\pi/2}\)).
  \item Pseudo-fermion heatbath (\(\phi \propto r(\m)^{-1}\xi\), where
    \(P(\xi)\propto e^{-\xi^*\xi}\))
  \item MD trajectory with \(\tau/\delta\tau\) steps
  \end{itemize}
\item Metropolis Acceptance Test \(\pacc = \mbox{min}
  (1,e^{-\dH})\).
\end{itemize}
Evaluating the force of this fermion action would result in a double
inversion because of the presence of the square of the rational
function.  To get around this a different rational approximation is
used for the MD: \(\bar{r} \approx \m^{-\alpha} \approx r^2\).  In
this form, the pseudofermion force is just a sum of HMC like terms
\begin{equation}
S'_{\rm pf} = -\sum_{i=1}^{\bar{m}} \bar{\alpha}_i
\phi^{\dagger}({\mathcal M}+\bar{\beta}_i)^{-1}{\mathcal
  M}'({\mathcal M}+\bar{\beta}_i)^{-1}\phi.
\label{eq:rhmc-force}
\end{equation}
The cost of the algorithm is almost identical to that of HMC, in that
there is one fermion inversion per force evaluation (there is one
extra inversion for the heatbath evaluation at the beginning of each
trajectory).

It is also perhaps worth mentioning that despite the fact that the
pseudofermion is held fixed throughout the MD trajectory, one cannot
use a chronological inverter \cite{Brower:1995vx} to reduce the number
of inversion iterations because of the requirement of the multi-shift
solver that the initial guess must be zero.  This however, represents
no handicap in practice since the chronological inverter isn't
applicable to the \(R\) algorithm or PHMC either.

\section{Finite Temperature}

\noindent The ideal testing ground for the RHMC algorithm is in the
regime of finite temperature QCD.  Most calculations here are done
using staggered fermions, with \(\Nf=2+1,3\).  Traditionally the \(R\)
algorithm has been used and so previous results have had finite
stepsize errors.  It is of course a valid question to ask whether
these previous results used a small enough stepsize to ensure that the
stepsize errors are negligible: is \(\dt^R\sim\frac{2}{3}\ml\) a valid
prescription for choosing the stepsize?

\begin{figure}
\begin{center}
\includegraphics[height=5cm]{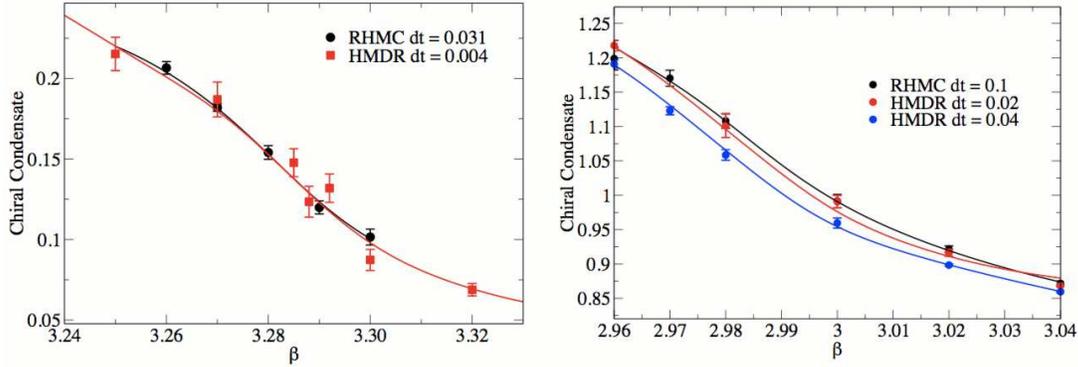}
\caption{A comparison of the chiral condensate between the \(R\) and
RHMC algorithms using P4 improved staggered fermions (\(V=8^34\),
\(\Nf=3\), left panel \(m=0.01\), right panel \(m=0.1\))
\cite{Cheng:2006fy}.}
\label{fig:p4rhmc}
\end{center}
\end{figure}

The RBC-Bielefeld collaboration use P4 improved staggered fermions for
their finite temperature calculations.  They have carried out various
comparisons between results generated using the \(R\) and RHMC
algorithms at different values of the fermion mass
\cite{Cheng:2006fy}.  In figure \ref{fig:p4rhmc} we can see
comparisons of the chiral condensate at two different fermion masses.
It was found for both fermion masses that a stepsize considerably
smaller than the \(\dt^R\sim\frac{2}{3}\ml\) was required to achieve
agreement between the two algorithms.  In all cases they found that
RHMC greatly accelerated the process of generating gauge fields by use
a much larger stepsize (i.e. \(\dt^{RHMC}\gg\dt^R\)), this being
especially true as the fermion mass is brought to zero.  As a result
of this study, all current gauge field production now exclusively uses
RHMC.

\begin{figure}
\begin{center}
\includegraphics[height=5cm]{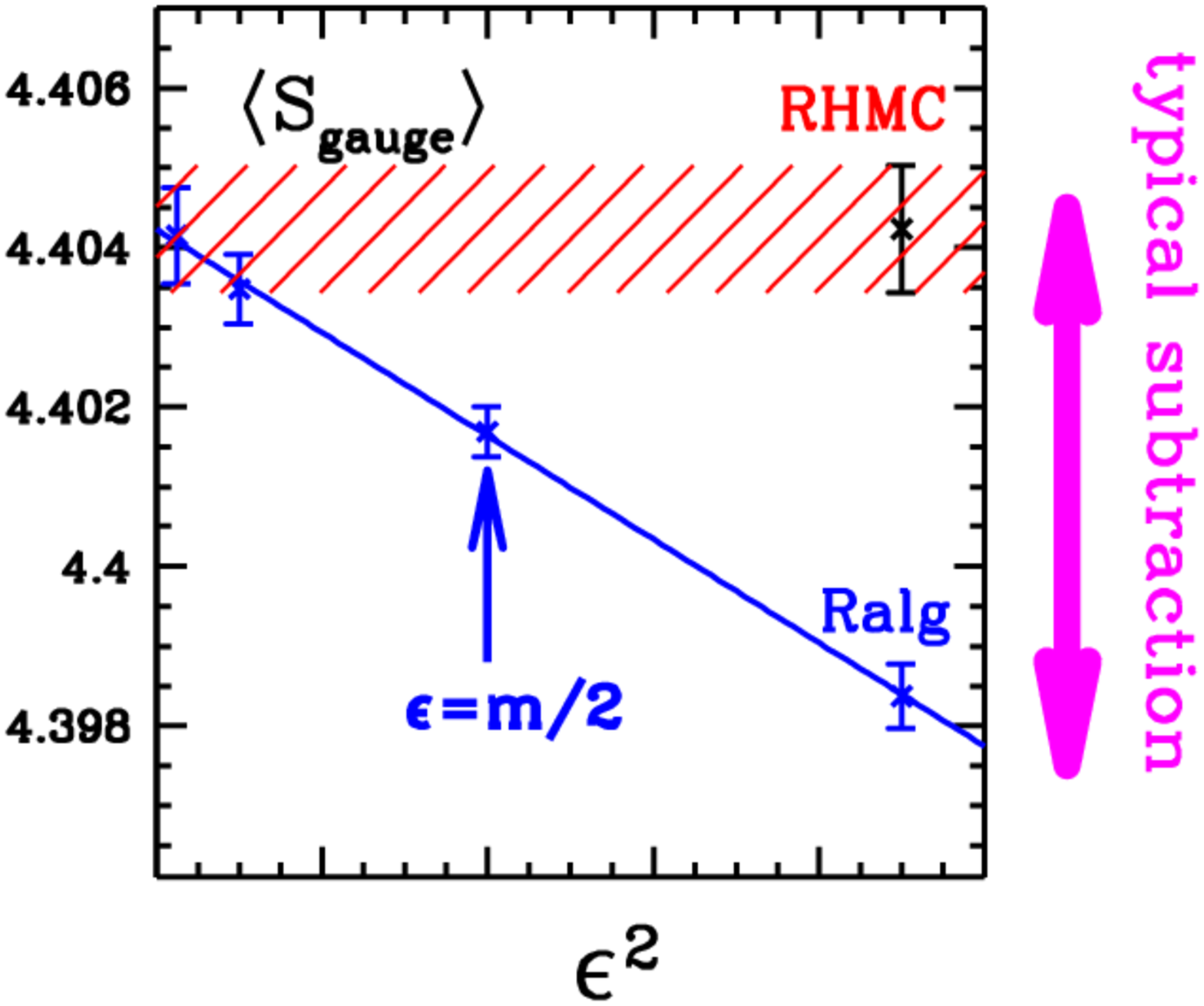}
\hspace{4mm}
\includegraphics[height=5cm]{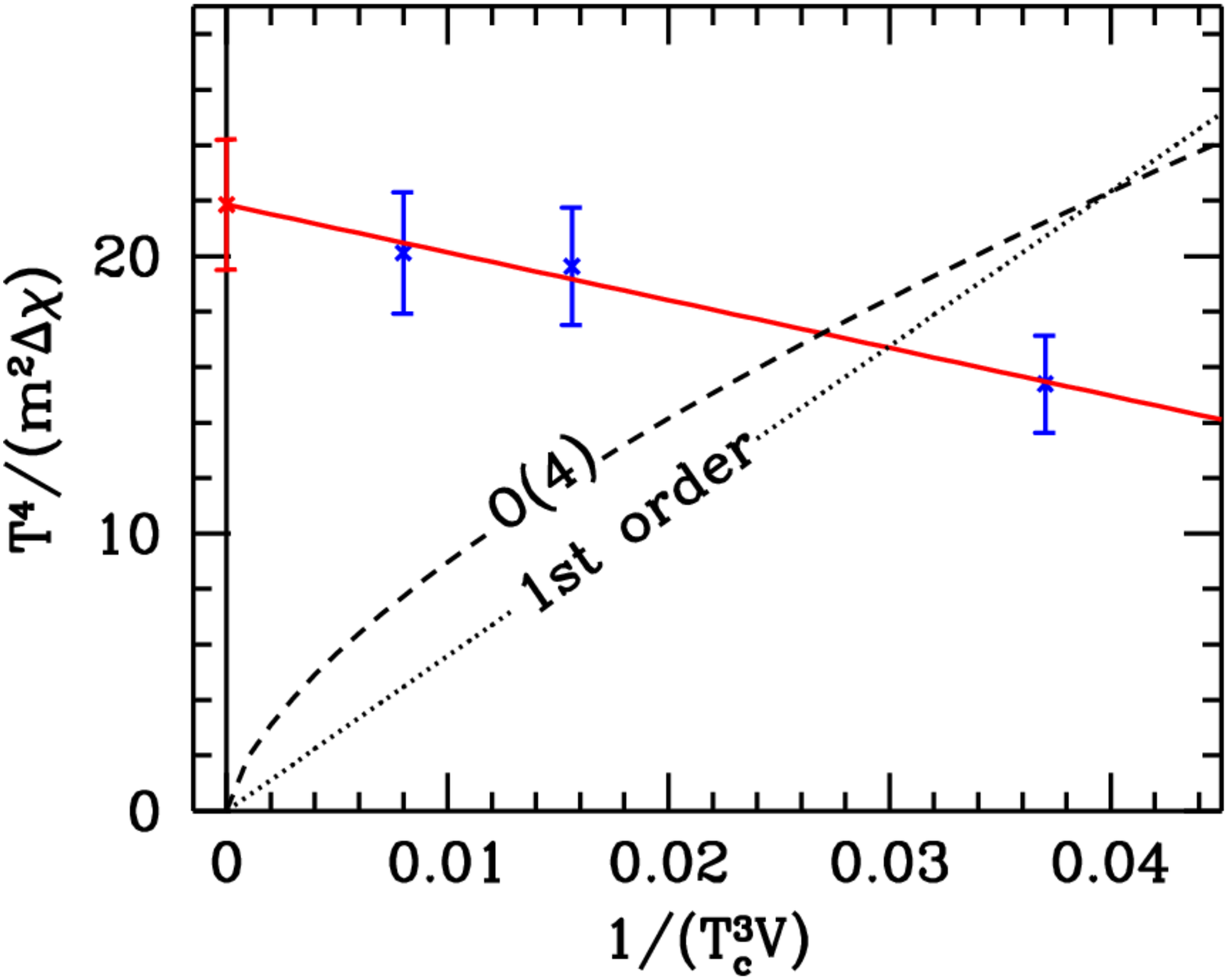}
\caption{Left panel: a plot of the expectation of the gauge of the
gauge action: the deviation of \(R\) and RHMC at
\(\dt^R\sim\frac{2}{3}\ml\) is comparable to the size of a typical
subtraction when calculating the equation of state (\(\epsilon=\dt\)
in the plot).  Right panel: generated using RHMC demonstrating that
the nature of the transition is a smooth crossover (stout improved
staggered fermions, \(V=16^4\), \(\Nf=2+1\), left panel:
\(m_\pi=\)320 MeV, right panel: \(m_\pi=\)140 MeV)
\cite{Wuppertal-Julich}.}
\label{fig:stoutrhmc}
\end{center}
\end{figure}

When studying the QCD equation of state and the order of the chiral
transition one has to subtract zero temperature observables from
finite temperature observables.  Using stout smeared staggered
fermions to study this behaviour, the Wuppertal-Budapest group have
found that the finite stepsize error using \(R\) algorithm is the same
order of magnitude as the typical subtraction (see figure
\ref{fig:stoutrhmc} left panel).  This makes the use of an exact
algorithm mandatory when making such calculations.

The gain in performance from the RHMC algorithm has allowed, for the
first time, finite temperature QCD to be studied using physical
fermion masses \cite{Aoki:2005vt, Aoki:2006br}.  This study has
revealed that the nature of the transition is a smooth crossover (see
figure \ref{fig:stoutrhmc} right panel).

\begin{figure}
\begin{center}
\includegraphics[height=5cm]{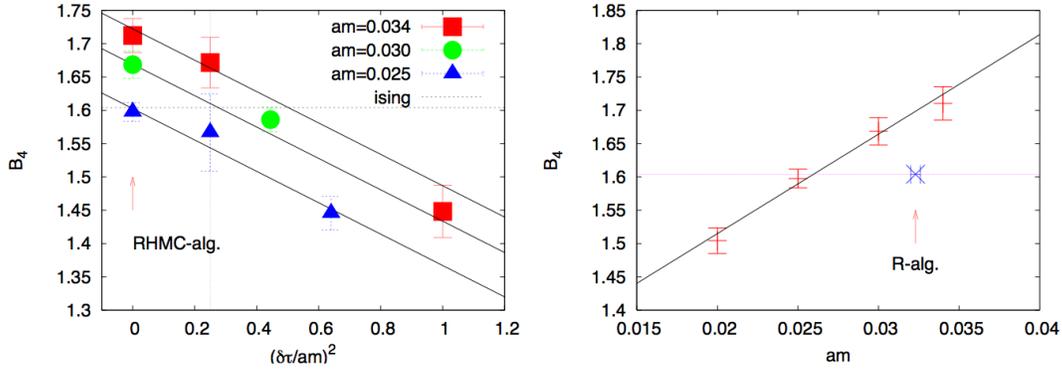}
\caption{Left panel: a plot demonstrating the increase in Binder
cumulant as \(\dt^R\) is reduced, tending towards the RHMC result.
Right panel: the shift in the critical quark mass when switching from
\(R\) (cross) to RHMC (line) (\naive\ Staggered Fermions, \(\Nf=3\),
\(V=8^34\)) \cite{deForcrand:2006pv}.}
\label{fig:stagrhmc}
\end{center}
\end{figure}

A thorough study of the stepsize errors of the \(R\) algorithm has
been carried out by de Forcrand and Philipsen
\cite{deForcrand:2006pv}.  They have found that the finite stepsize
error causes the Binder cumulant to increase (see figure
\ref{fig:stagrhmc} right panel) similar to other previous studies
\cite{Kogut:2005qg}, and again find significant cost reduction from
switching from \(R\) to RHMC.  More significantly at
\(\dt^R=\frac{1}{2}\ml\), the critical fermion mass is overestimated
by 33\% compared to the exact result.  This corresponds to a 25\%
increase in the renormalised fermion mass.  The finite stepsize error
is present in renormalised quantities, this is in direct contradiction
of the statement in \cite{Davies:1985ad}.

The only possible conclusion that can be made from these finite
temperature studies is that an exact algorithm isn't just desirable
from a cost point of view, but it is mandatory to ensure that the
physics obtained is free from systematic error.

\section{Multiple Pseudofermions}
\label{sec:multiple}
\subsection{Mass Preconditioning}
\noindent A large gain in performance has been found in the technique
of mass preconditioning the fermion determinant
\cite{Hasenbusch:2001ne}
\[
\det (M^\dagger M) = \det (\hat{M}^\dagger \hat{M}) \det \left(
\hat{M}^{-1} (M^\dagger M) \hat{M}^{-\dagger} \right)
\]
with \(m(\hat{M})>m(M)\).  This product of determinants can then
represented using two pseudofermions, with kernels \((\hat{M}^\dagger
\hat{M})^{-1}\) and \( \hat{M} (M^\dagger M)^{-1} \hat{M}^\dagger\)
respectively.  Using multiple pseudofermions to represent the
determinant is beneficial because it reduces the fluctuations in the
fermion force through better sampling of the Gaussian integral.
Compared to the standard formulation there is a relatively small
increase in \naive\ cost because of the need to invert the extra
operator in the second bilinear, but the use of multiple
pseudofermions allows for a large increase in the stepsize: the end
result being a net gain in performance.

The initial strategy for mass preconditioning was to tune the dummy
operator's mass parameter such that
\[
\kappa(\hat{M}^\dagger\hat{M}) \approx \kappa( \hat{M} (M^\dagger
M)^{-1} \hat{M}^\dagger),
\]
i.e., tune such that the condition number of the two kernels are equal
\cite{Hasenbusch:2002ai}.  This can gain up to a factor 2 improvement
though the technique can be extended to introduce more pseudofermions
potentially leading to larger performance gains.

\subsection{Multi-timescale Mass Preconditioning}
\begin{figure}
\begin{center}
\includegraphics[height=5cm]{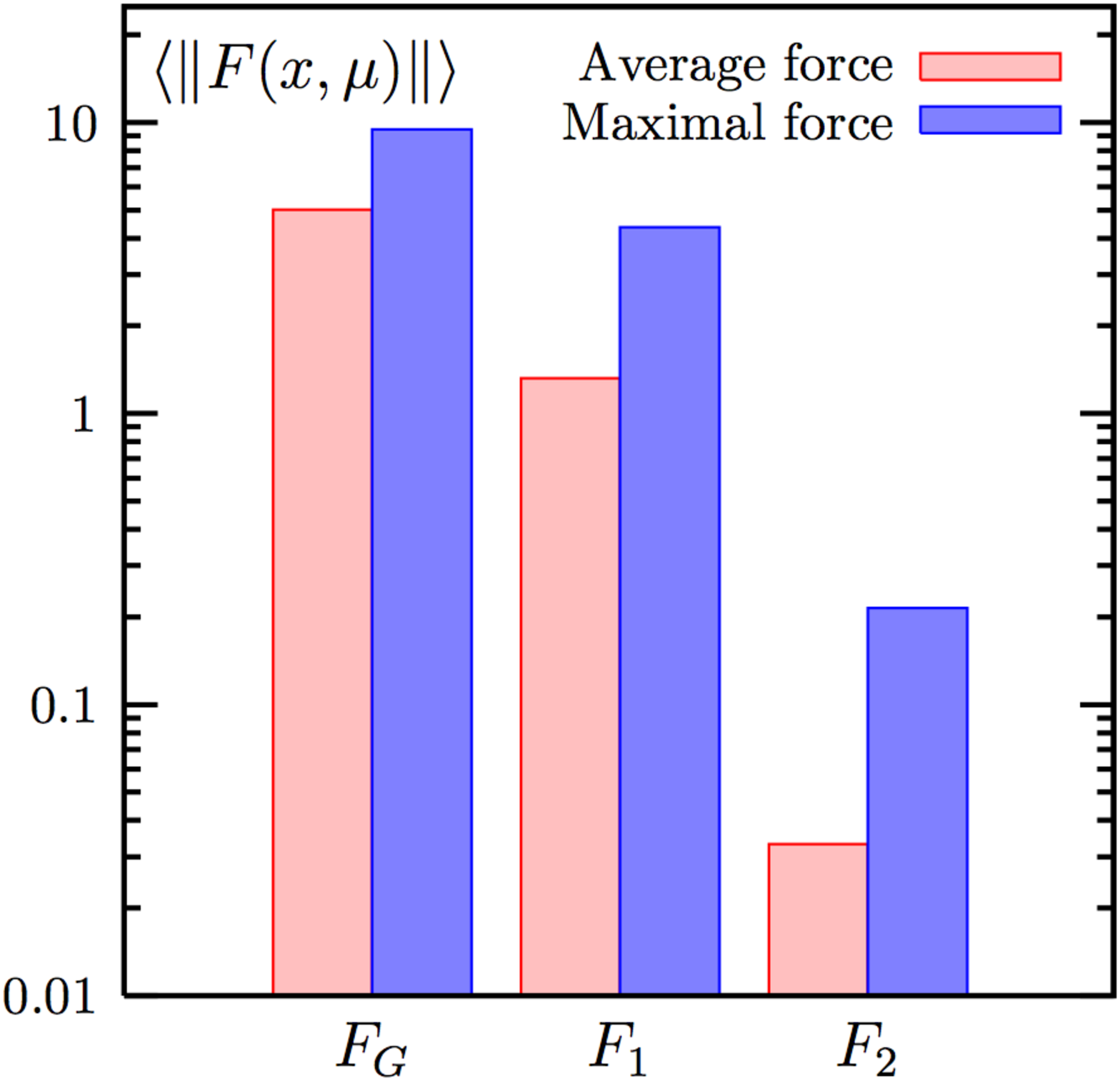}
\hspace{4mm}
\includegraphics[height=5cm]{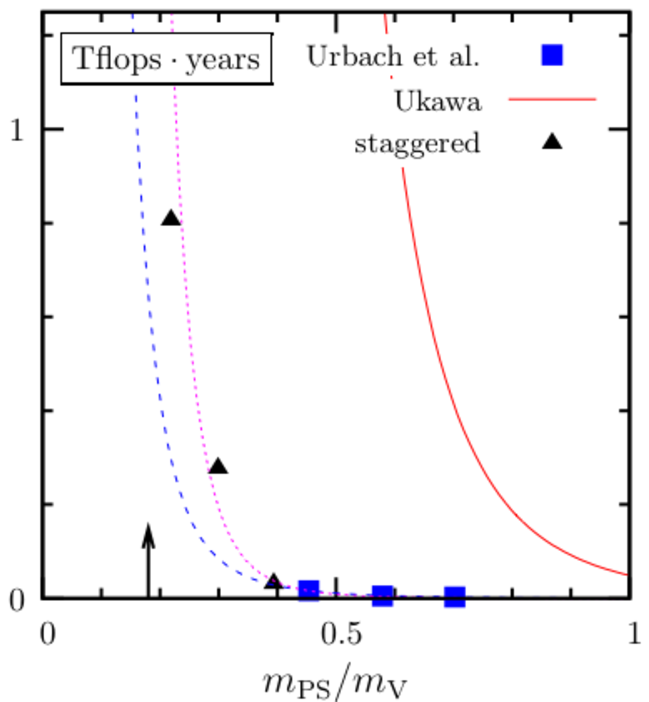}
\caption{Left panel: the magnitude of the forces present in a
multi-timescale mass preconditioned simulation: the most expensive
force (\(F_2\)) contributes the least to the total force and so can be
included less often in the integration.  Right panel: cost to generate
1000 independent configurations for Wilson fermions using mass
preconditioning (blue squares) and conventional HMC (red line)
compared to staggered fermions with the \(R\) algorithm (Wilson
fermions, \(V=24^332\), \(\Nf=2\), \(\beta=5.6\))
\cite{Urbach:2005ji}.}
\label{fig:urbach}
\end{center}
\end{figure}

\noindent An alternative to the strategy suggested above is to tune
the additional dummy operators such that the most expensive force
contributes the least magnitude to the total force
\cite{AliKhan:2003br, Urbach:2005ji} (see figure \ref{fig:urbach} left
panel).  This strategy then leads to large gains when used in
combination with a multi-timescale (nested) integrator
\cite{Sexton:1992nu}: as the fermion mass is brought to zero, the gain
is found to be around a factor 10 over the basic HMC algorithm at
current light fermion parameters (see figure \ref{fig:urbach} right
panel).
It is worth mentioning that this achieves at least the performance of
the domain decomposition \cite{Luscher:2005rx}, but is much easier to
implement.

\subsection{Multiple Pseudofermions with RHMC}
\noindent We can trivially rewrite fermion determinant
\[
  \det\m  = [\det\m^{1/n}]^n \nonumber \propto \prod_{j=1}^n d\phi_j\, d\phi^\dagger_j\,
    \exp{\left(-\phi^\dagger_j\m^{-1/n}\phi_j \right)}, \nonumber
\]
this is the so called \(n^{th}\) root trick, and this action can
easily be simulated using RHMC \cite{Clark:2004cq,Clark:2006fx}.  Like
the multiple pseudofermion technique of Hasenbusch, this alternative
method leads to a speedup through an increase in stepsize.  This
method has the added virtue that there are no dummy mass parameters to
tune, so it easy to extend to arbitrary \(n\).  For this approach,
since all fermion inversions involve the same light kernel a single
timescale is used for the fermions.

\begin{figure}
\begin{center}
\hfill
\begin{minipage}[t]{0.45\textwidth}
\includegraphics[height=5cm]{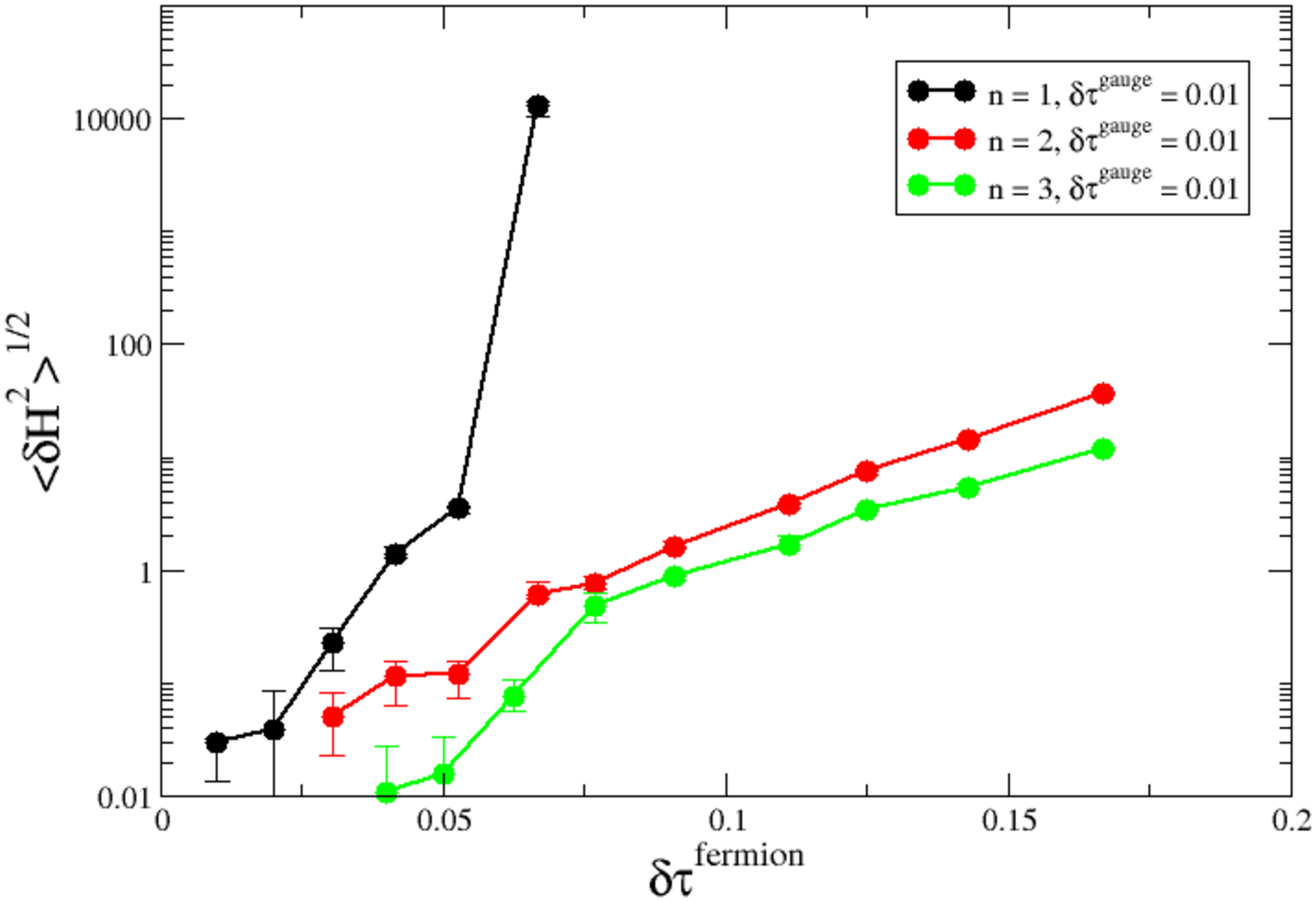}
\caption{A plot of magnitude of \(\dH\) versus the stepsize for
\(n=1,2,3\) pseudofermions with RHMC.  The instability is triggered
only for \(n=1\) (staggered fermions, \(V=16^4\), \(\beta=5.6\),
\(\Nf=2\), \(m=0.005\)) \cite{Clark:2006fx}.}
\label{fig:dh_dt}
\end{minipage}
\hfill
\begin{minipage}[t]{0.45\textwidth}
\includegraphics[height=5cm]{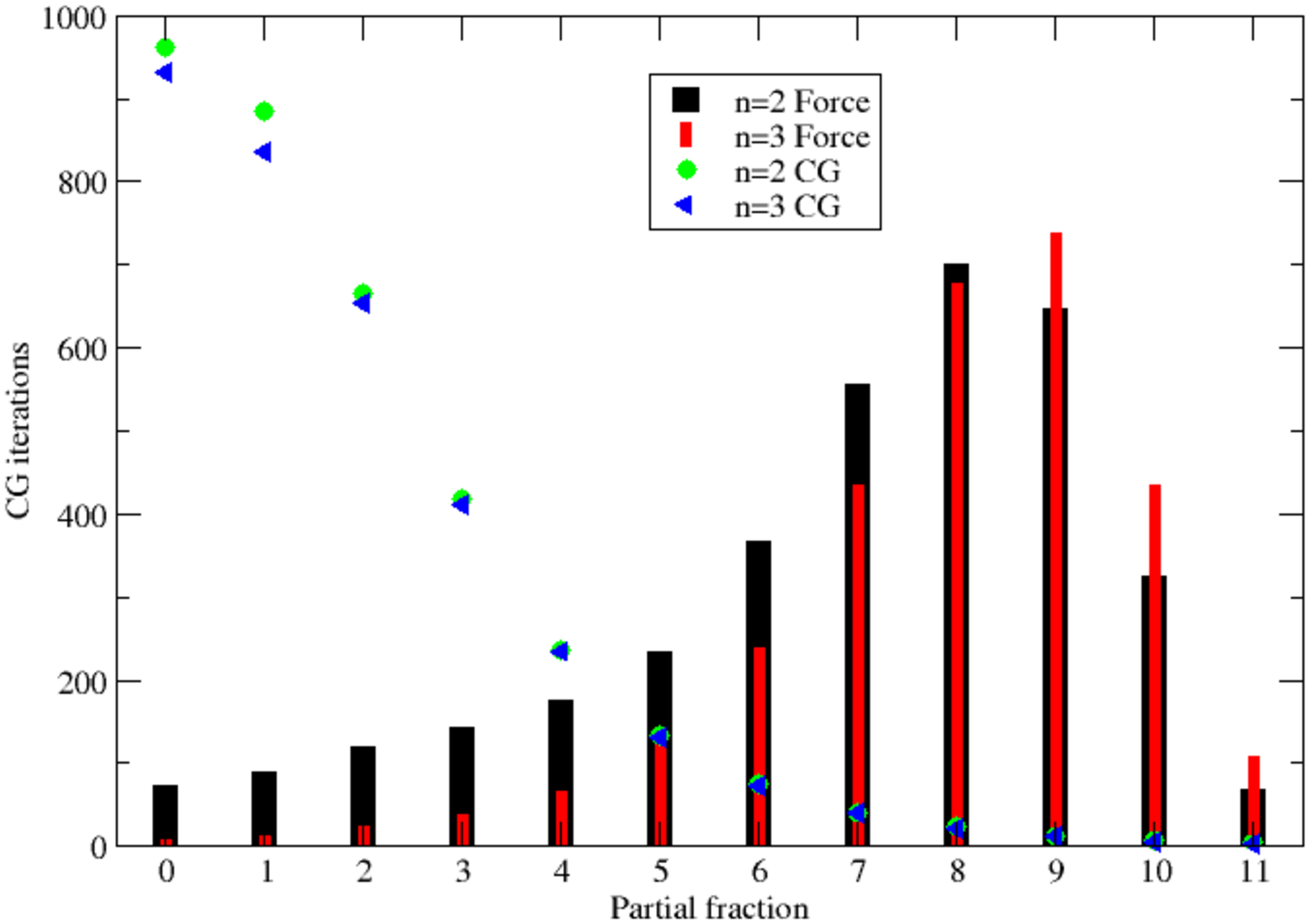}
\caption{A plot of the number of conjugate gradient (CG) iterations
required per pole to converge in the multi-shift solver and of the
relative force contribution arising from each for \(n=2,3\)
pseudofermions (Wilson fermions \(\Nf=2\), \(\beta=5.6\),
\(\kappa=0.15800\), \(V=24^3\times32\)) \cite{Clark:2006fx}.}
\label{fig:wilson-force}
\end{minipage}
\hfill
\end{center}
\end{figure}

Before we answer the question as to which of the multiple
pseudofermion methods leads to the best performance, it is interesting
to look at the question of integrator instability.  On figure
\ref{fig:dh_dt} we have plotted the magnitude of \(\dH\) as a
function of stepsize.  With \(n=1\) pseudofermions we can see that a
point is reached where by the instability in the integrator is
triggered, and there is catastrophic breakdown of Hamiltonian
conservation.  This instability is caused by the light fermion modes
\cite{Joo:2000dh} which have magnitude \(\sim O(\frac{1}{\ml^2})\).
It is the presence of the instability that results in higher order
integrators being detrimental for HMC: this is because such
integrators are constructed out of leapfrog sub-steps which are longer
than the overall step and are of course susceptible to the
instability.  What happens to this picture when multiple
pseudofermions are used?  On the same figure we have also included the
equivalent results for \(n=2,3\).  We can see that the improved
Hamiltonian conservation from using multiple-pseudofermions as
expected, but crucially the integrator instability has been removed.
This is because the low modes which cause the instability, which were
\(O(\frac{1}{\ml^2})\) are now \(O(\frac{1}{\ml^2})^{1/n}\).  Thus
since the limiting light modes have effectively been removed, we
expect to gain significantly from the use of higher order integrators:
this is indeed found to be the case, we have found a gain of a factor
2 switching from second order to fourth order integrators
\cite{Clark:2006fx}.  The switch to a higher order integrator is also
expected to lead to improved volume scaling, i.e., the expected
scaling of HMC with a second order integrator is \(V^{5/4}\), where as
with a fourth order integrator it is \(V^{9/8}\) \cite{Creutz:1989wt}.

\subsubsection{Partial Fraction Forces}

It was noted in equation \ref{eq:rhmc-force} that the forces are
merely a sum of HMC like force terms.  Each of these forces is
expected to have a different magnitude, and so contribute differing
amounts to the total fermion force.  Figure \ref{fig:wilson-force} is
plot of the break down of the magnitude of the force originating from
each partial fraction for \(n=2,3\) pseudofermions.  The somewhat
surprising result is that the terms which contribute the least to the
total fermion force are those which cost the most (compare with the
conjugate gradient (CG) iteration count).  The force distribution
arises primarily because of the nature of the rational coefficients.
We would hope that we can make the most of this observation, and
indeed we can: we loosen the CG tolerance on the less well conditioned
shifts, i.e., those which contribute the least to the fermion force.
This can strategy can gain up to a factor of 2 reduction in total CG
iterations per trajectory.  Of course the tolerance is not loosened
for the heatbath and Metropolis evaluation to avoid any bias being
introduced.  As \(n\) is increased this effect is enhanced as can be
seen on the plot.

\subsection{ Who's the fastest of them all? (\(\Nf=2\) Wilson)}

We now compare the two multiple pseudofermion techniques presented so
far in this paper: mass preconditioning and \(\rmsup{n}{th}\) roots.
To do this comparison the parameters and mass preconditioning results
from \cite{Urbach:2005ji} were used.  The measure of efficiency that
was adopted in this work was used here: the product of the
autocorrelation length of the plaquette \(\tint{plaq}\) with the
number of matrix vector products per trajectory \(\Nmv\).

\begin{table}
  \begin{center}
  \begin{tabular}{|l|c|c|c|} \hline
    & \multicolumn{3}{|c|}{\(\tint{plaq}.\Nmv .10^{-4}\)} \\ \hline
    \(\kappa\) & RHMC \cite{Clark:2006fx} & Mass preconditioning
    \cite{Urbach:2005ji} & HMC \cite{Orth:2005kq} \\ \hline
    0.15750    & 9.6 & 9.0 & 19.1 \\ \hline
    0.15800    & 19.1* & 17.4 & 128 \\ \hline
    0.15825    & 52.5* & 56.5 & - \\ \hline
  \end{tabular}
\caption{A comparison of the costs of RHMC, mass preconditioning and
  plain HMC (Wilson fermions, \(V=24^3.32\), \(\beta=5.6\), *Using
  4MN5 fourth order integrator \cite{Takaishi:2005tz})}
\label{table:wilson-results}
\end{center}
\end{table}

As can be seen in table \ref{table:wilson-results}, the cost of the
two algorithms is extremely similar and there is little to chose
between the two.  As is always the case, it would be interesting the
compare the algorithms' relative performance as the chiral limit is
taken.  Given that the single timescale \(\rmsup{n}{th}\) root
formulation is using a higher order integrator, we would expect
superior volume scaling.  It would also be interesting to investigate
whether the use of higher order integrators brings the same benefit to
mass preconditioning.

\section{Algorithmic Improvement with \(\Nf=2+1\)}
\subsection{Domain Wall Fermions}

\begin{figure}
\begin{center}
\includegraphics[height=5cm]{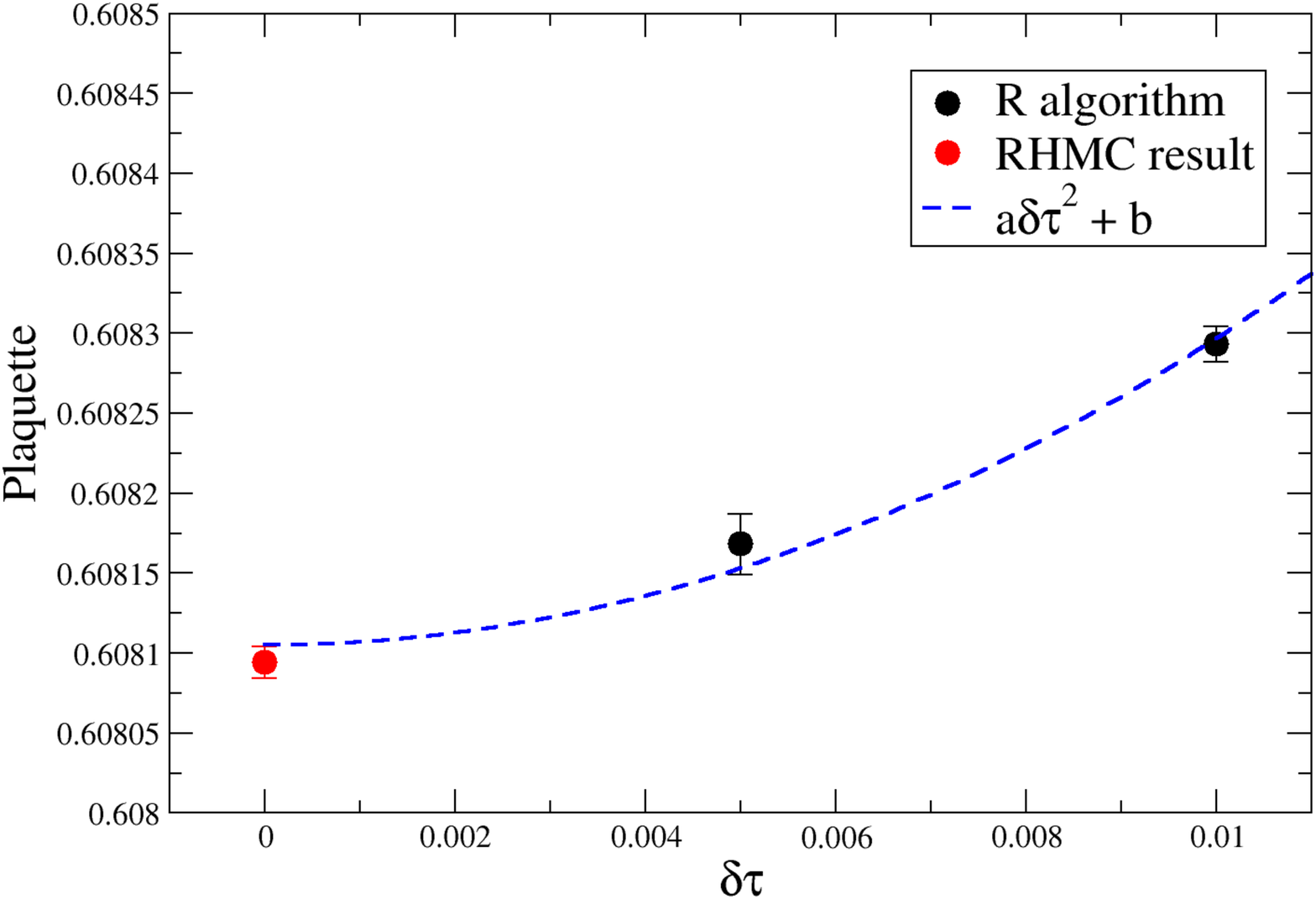}
\includegraphics[height=5cm]{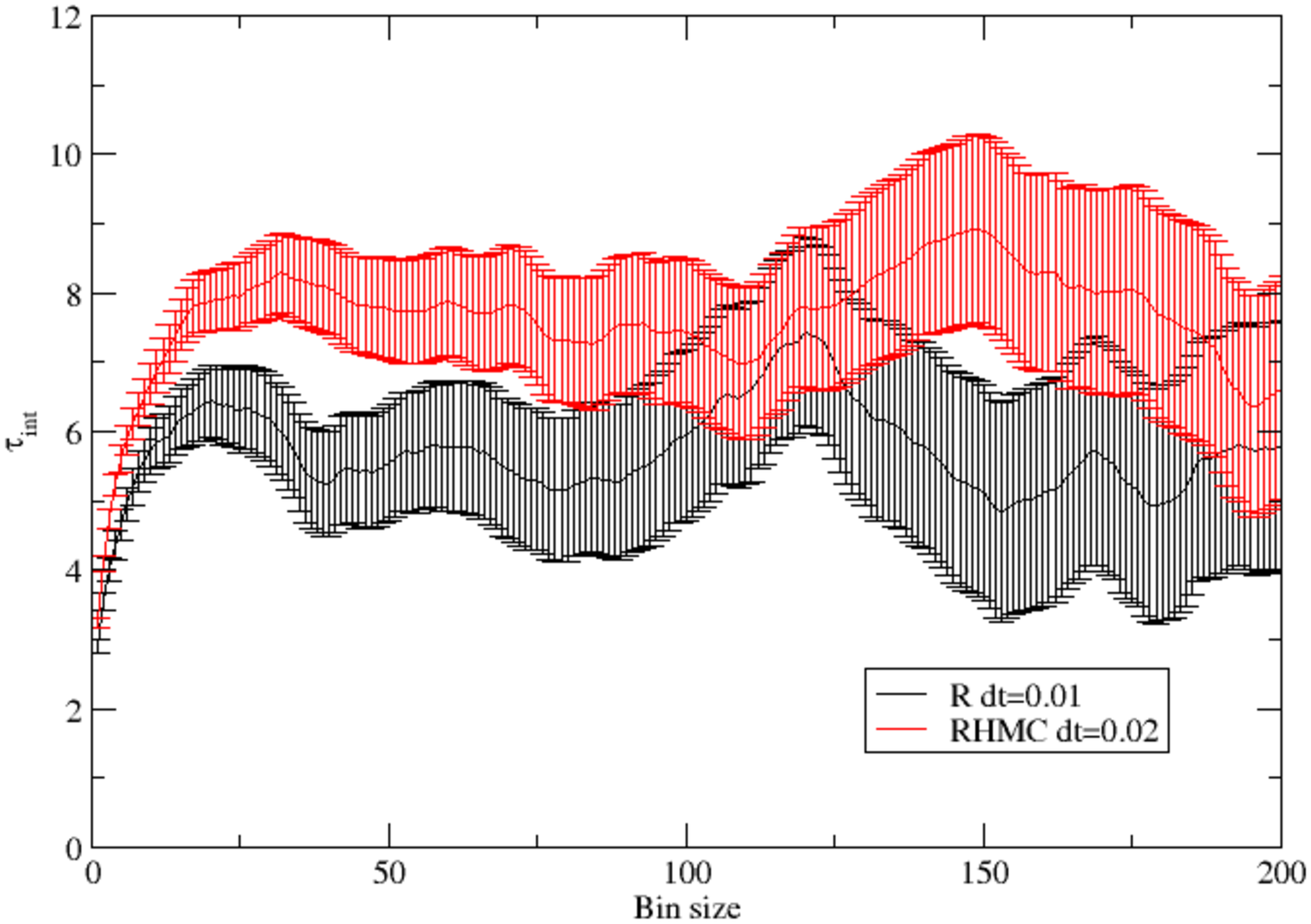}
\caption{Left panel: \(R\) and RHMC plaquette values, the
extrapolation to zero stepsize is clearly consistent with the RHMC
result.  Right panel: The plaquette autocorrelation length of \(R\)
and RHMC (domain wall fermions, \(V=16^3\times32\times8\),
\(\Nf=2+1\),\(\beta=0.72\), \(\ml=0.02\), \(\ms=0.04\),
\(\dt^{RHMC}=0.02\)) \cite{Clark:2005sq}.}
\label{fig:dwf}
\end{center}
\end{figure}

\noindent The first large production runs of RHMC were that of the
\(\Nf=2+1\) program by the RBC-UKQCD collaboration
\cite{Clark:2005sq,Antonio:2005yh}.  As a first test of the algorithm,
the standard RHMC algorithm was compared against the \(R\) algorithm.
Figure \ref{fig:dwf} is taken from this initial study, where in the
left panel the \(R\) algorithm's plaquette measurement is seen to
extrapolate to the RHMC result at zero stepsize.  A comparison of the
algorithms' autocorrelation length revealed the expected difference
due to the non-unit acceptance rate of RHMC.  With this initial study
completed, it was decided to use RHMC for all subsequent running.

A somewhat obvious observation can be made when simulating a
\(\Nf=2+1\) theory.  For the current case of domain wall fermions, we
write down the 2+1 flavour determinant
\[
\left(\frac{\det \mf{l} }{\det \mf{pv} }\right)\left(\frac{\det \mf{s} }{\det \mf{pv} }\right)^{1/2} = 
\left(\frac{\det \mf{l} }{\det \mf{s} }\right)\left(\frac{\det \mf{s} }{\det \mf{pv} }\right)^{3/2},
\]
where in the right hand side we have trivially rewritten the
determinant product.  What this suggests is that we should mass
precondition the light quark by the strange quark.  Unlike the two
flavour case, since we already have the heavier fermion present, there
is no extra overhead from doing mass preconditioning.  As is suggested
in \cite{Urbach:2005ji} we can then use a multiple timescale
integration scheme (gauge, triple strange, preconditioned light pair)
to make maximum use of the mass preconditioning.  This strategy is
found to work extremely well, and leads to around a large reduction in
CG cost \cite{Mawhinney:2006}.  It is interesting to note that in this
arrangement, the light quark inversions take up around 10\% of the
total CG cost, and thus the mass dependence of the cost has
drastically reduced since generally the strange quark is held fixed.

\subsection{AsqTad Fermions}

\noindent Apart from finite temperature QCD calculations the other
obvious target for the RHMC algorithm is the large scale \asqtad\
calculations that are done by the MILC collaboration.  A significant
cost reduction here would correspond to a saving of many Teraflop
years of computer time.

\begin{figure}
\begin{center}
\hfill
\begin{minipage}[t]{0.45\textwidth}
\includegraphics[height=5cm]{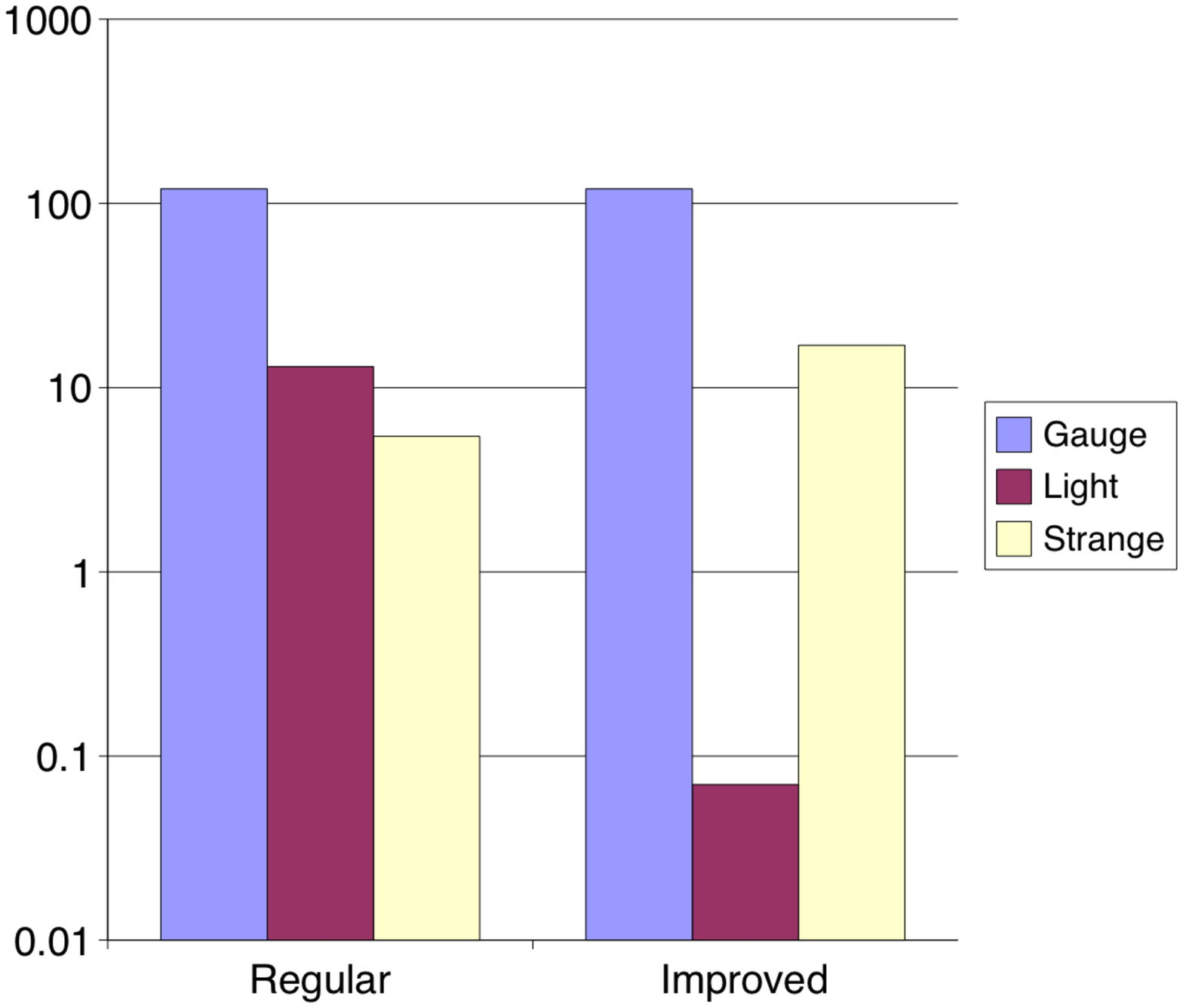}
\caption{Comparison of the relative force magnitudes before and after
preconditioning the light quark by the strange (\asqtad\ fermions,
(\(V=4^4\), \(\Nf = 2+1\), \(\ml=0.01\), \(\ms=0.05\), \(\beta=6.76\))
\cite{Clark:2006asqtad}.}
\label{fig:asqtad-force}
\end{minipage}
\hfill
\begin{minipage}[t]{0.45\textwidth}
  \includegraphics[height=5cm]{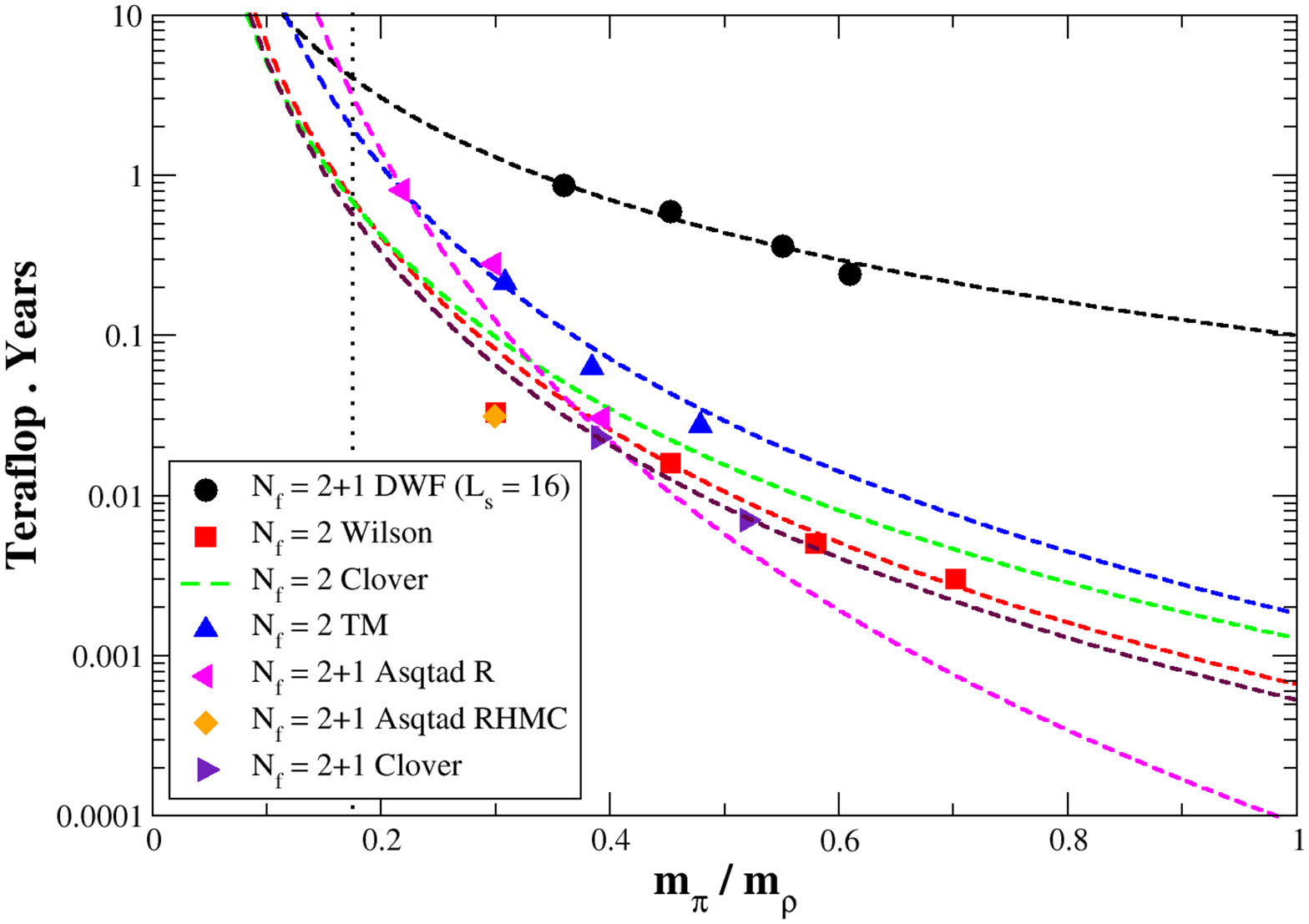}
  \caption{An updated Berlin wall plot comparing the costs of various fermion formulations.}
  \label{fig:berlin}
\end{minipage}
\hfill
\end{center}
\end{figure}

Following on from the idea of mass preconditioning using the strange
quark developed in the previous section, here we do the same but with
\asqtad\ fermions,
\[
\det(\Ml)^{\frac{1}{2}}\det(\Ms)^{\frac{1}{4}} =
\left(\frac{\det(\Ml)}{\det(\Ms)}\right)^{\frac{1}{2}} \, \det(\Ms)^{\frac{3}{4}}.
\]
With staggered kernels, the mass is simply a multiple of the identity,
this leads to a particularly simple form of fermion action since we
can represent the ratio function by a single rational approximation
\cite{Clark:2006asqtad}
\begin{eqnarray}
  \Sf & = & \phil^\dagger \left(\frac{\Ms}{\Ml}\right)^{\frac{1}{2}}\phil +
  \phis^\dagger \Ms^{-\frac{3}{4}} \phis \nonumber \\
  & = & \phil^\dagger \left(\frac{\Ml + \delta m^2}{\Ml}\right)^{\frac{1}{2}}\phil +
  \phis^\dagger \Ms^{-\frac{3}{4}} \phis \nonumber \\
  & = & \phil^\dagger r^2(\Ml) \phil + \phis^\dagger r^2(\Ms) \phis \nonumber.
\end{eqnarray}
In figure \ref{fig:asqtad-force} we have plotted the breakdown of the
force magnitudes before and after the mass preconditioning.  The
effect from the preconditioning is apparent in the plot: the light
quark contribution is reduced by an order of magnitude and the strange
quark increases by a factor of 3, as would be expected.
Again, we can take advantage of this through the use of a multiple
timescale integrator.  The mass preconditioning allows for a vast
reduction in the CG cost of producing \asqtad\ ensembles, but this
does not correspond to a likewise reduction in the total operation
count: the triple strangle is the dominant cost as expected but this
cost does not come from CG iterations, rather from the expensive
operator derivative calculation that is required with \asqtad\
fermions \cite{Clark:2004cp}.  Thus any further mass preconditioning
is detrimental to performance because although it may reduce the CG
count, it introduces more operations since each timescale requires a
separate derivative calculation.  The solution to this problem is to
use the \(\rmsup{n}{th}\) root method the triple strange.  This
introduces more pseudofermions all on a single timescale, allowing
each field to reuse the same derivative calculation, greatly reducing
the cost, while increasing the stepsize of the triple strange
timescale.  Thus the optimum solution uses mass preconditioning and
the \(n^{th}\) root trick.  When this improved technique is tested at
current MILC parameters (\(V=24^3\times 64\), \(\beta=6.76\),
\(\ml=0.005\), \(\ms=0.05\)) against the \(R\) algorithm it is found
to lead to a factor 7-8 reduction in the operation count per
trajectory, and of course the results generated are exact
\cite{Clark:2006asqtad}.

\section{Berlin Wall Plot}

\noindent The standard method for demonstrating the mass scaling cost
of lattice fermion formulations is to use the so called Berlin wall
plot.  On the x-axis is plotted the ratio of the pion and rho masses
and on the y-axis is the cost in Teraflop years to generate 1000
independent gauge configurations.  With all of the algorithmic
improvements that have been applied to QCD in recent years, an updated
Berlin wall plot is necessary to compare the costs of all the fermion
formulations.  For this plot, the following formulations (and
algorithms) have been included: \(\Nf=2+1\) DWF RHMC \cite{RBC-UKQCD},
\(\Nf=2\) mass preconditioned Wilson \cite{Urbach:2005ji}, \(\Nf=2\)
mass preconditioned clover \cite{QCDSF}, \(\Nf=2+1\) mass
preconditioned clover + RHMC \cite{Wuppertal-Julich}, \(\Nf=2\) mass
preconditioned twisted mass \cite{ETM}, \(\Nf=2+1\) \asqtad\ R
\cite{MILC} and \(\Nf=2+1\) \asqtad\ RHMC \cite{Clark:2006asqtad}.

Since these various results have been run at different lattice spacing
and volumes, the cost data must be scaled to be consistent.  Hence,
all data has been scaled to \(V=24^3\times40\), \(a=0.08\).  A word of
warning about this box size and lattice spacing: this box is far too
small and coarse to realistically represent the physical point, so the
scaling behaviour as the physical point is approached should be taken
with the requisite amount of salt.

The updated Berlin wall plot can be seen in figure \ref{fig:berlin}.
The cost of using domain wall fermions is around the expected factor
of \(\rmsub{L}{s}\) times more expensive than the Wilson formulations.
All Wilson formulations would appear to have a similar cost and
scaling.  What is also evident from the plot is that before switching
to RHMC, the cost of using \asqtad\ fermions is significantly more
than that of mass preconditioned Wilson fermions.  When mass
preconditioned RHMC is used the cost of \asqtad\ fermions seems to
fall down to the level of Wilson fermions.  The main point to be drawn
from this plot though is that the mass dependences is much weaker for
the improved algorithms than is suggested in \cite{Ukawa:2002pc}.

\section{Summary}

In this paper we have described the exact Rational Hybrid Monte Carlo
algorithm and explained how it can be used for fermion theories where
a non-integer power of the determinant is required.  

In the regime of finite temperature QCD studies, where having an exact
algorithm is crucial, the RHMC algorithm was shown to be far superior
to the R algorithm.  Indeed, since using an exact algorithm is cheaper
than an inexact one, there appears to be no need for the continued use
of the R algorithm.

With the use of the \(\nth\) root method of pseudofermions, RHMC is
also an effective algorithm for theories with an integer power of the
determinant.  It was shown that the use of multiple pseudofermions
removes the integrator instability, this mean that the use of higher
order integrators can be very beneficial.  Both this method, and that
of mass preconditioning lead to large gains in performance compared to
conventional HMC.  These techniques can even be combined, and the use
of RHMC allows one to use the strange quark to precondition the light
quarks if \(\Nf=2+1\) is considered.

The use of improved algorithms has greatly reduced the mass dependence
of the cost of generating gauge configurations.  Although we are not
yet at a stage where we can simulate using physical quark masses, the
possibility of doing so does not now lie that far out of reach.

\section{Acknowledgments}

The author would like to thank the following people for supplying
data, figures and thoughts used in this work: Michael Cheng, Zoltan
Fodor, Steve Gottlieb, Karl Jansen, Tony Kennedy, Bob Mawhinney,
Gerrit Schierholz, Kalman Szab\'{o} and Carsten Urbach.




\end{document}